\documentclass[a4paper]{article}
\usepackage{graphicx}

\title{\bf Monte Carlo Algorithm for Least Dependent Non-Negative Mixture Decomposition}
\author{
{\bf Sergey A. Astakhov}, {\bf Harald St\"ogbauer}\\
{\bf Alexander Kraskov}, {\bf Peter Grassberger}
\\\\
John von Neumann Institute for Computing, \\ Forschungszentrum
J\"ulich, D-52425, J\"ulich, Germany, \\ and \\Division of Biology, California Institute of Technology, \\
California 91125, USA
\\\\
}

\begin{document}

\maketitle

\begin{abstract}
 We propose a simulated annealing algorithm (called SNICA for ``stochastic non-negative independent
 component analysis") for blind decomposition of linear mixtures of
 non-negative sources with non-negative coefficients.
 The de-mixing is based on a Metropolis type Monte Carlo search for least
 dependent components, with the mutual information between recovered components as a cost function and
 their non-negativity as a hard constraint. Elementary moves are shears in two-dimensional
 subspaces and rotations in three-dimensional subspaces. The algorithm is geared
 at decomposing signals whose probability densities
 peak at zero, the case typical in analytical spectroscopy and
 multivariate curve resolution. The decomposition performance on large samples of synthetic mixtures
 and experimental data is much better than that of
 traditional blind source separation methods based on principal component
 analysis (MILCA, FastICA, RADICAL) and chemometrics techniques (SIMPLISMA, ALS, BTEM).\\\\\\
\end{abstract}


\section{Introduction}
\label{intro}

Decomposing linear mixtures (superpositions) into their components is a problem
occurring in many different branches of science, such as telecommunications,
seismology, image processing, and biomedical signal analysis ({\it e.g.}, EEG, ECG,
fMRI) \cite{ica}. Blind source separation (BSS), in
particular, deals with the case where neither the sources, nor the mixing matrix
are known, the only available data being the mixed signals \cite{bss}.
The standard approach to BSS is {\it independent component analysis} (ICA)
\cite{ica,bss,com,jut}. In ICA one typically first applies a principle
component transformation to ``prewhiten" the data, {\it i.e.} to transform the
covariance matrix into a unit matrix, and then uses some statistics going
beyond second order ({\it e.g.} mutual information or temporal correlations) to
minimize interdependencies between the reconstructed sources by a final
rotation.

A similar problem arises in analytical spectroscopy, where one wants to decompose
spectra of chemical mixtures into the contributions of individual components in
order to identify the emitting (or absorbing) mixture constituents and quantify
their abundances. Mixture decomposition problems of this sort are classified in
analytical chemistry as quantitative analysis of ``black'' multicomponent systems
\cite{manne}. Compared to other applications of BSS, here we have two additional
problems:

\begin{itemize}
\item Spectra of different chemically pure substances often show very large
overlaps. This is in conflict with the central assumption of ICA that the pure
sources (mixture components) are statistically independent. One can relax this
assumption partially (see, {\it e.g.}, Refs.~\cite{milca,smd}), although one
can hardly do without any assumption of this type. But, because of overlaps
between spectra of similar compounds, this becomes a serious obstacle in
spectral mixture decomposition.

\item In a vast majority of spectroscopic techniques, spectra are non-negative,
as are also the coefficients of the mixing matrix (concentrations). This gives
rise to constraints which are not easily taken into account in typical ICA
algorithms that heavily rely on linear algebra \cite{con1,con2}. This would not be problematic
by itself -- in other applications signals are often non-negative, too. But it
does pose a very serious problem, if, as is usually the case, there are
spectral regions where some of the sources have zero (or very small) intensity
\cite{plumbley}. In this case, prewhitening will in general produce spectra
violating the non-negativity constraints, and restoring them later will be far
from trivial.
\end{itemize}

For these reasons, this ``Multivariate Spectral Curve Resolution" (MCR) problem
\cite{pioneer} has led to a variety of chemometrics methods (for recent reviews
see Refs.~\cite{geladi,hopke,jiang,tauler,lavine}). Generally, these algorithms
resemble ICA in that they perform blind recovery of spectra and concentrations
from the spectra of mixtures only, using no or little {\it a priori} information
about mixture composition and/or about pure spectra. In view of the problems
mentioned above, some of these MCR methods are less ambitious than ICA and
estimate only feasible ranges for spectra and concentrations (see, for
instance, a recent Monte Carlo approach \cite{mc} and references therein).
Also, the MCR analysis is often facilitated by including some additional
information such as unimodality and closure (or mass balance) \cite{tauler}. The effect of
overlap between spectra can be reduced by estimating their dependencies not
from the spectra $S(f)$ directly, but from their second derivatives
$d^2S(f)/df^2$ \cite{smd,windig}. Another approach developed in chemometrics is
based on the identification of ``pure variables'' \cite{pv} for all mixture
components. A pure variable is a frequency (wavelength) at which only one of
the components contributes. The pure variables, thus, approximately mark the
regions where at least one of the spectral components is guaranteed to be
independent from all others. This idea is central
to several advanced chemometrics algorithms, {\it e.g.}, KSFA \cite{ksfa}, SIMPLISMA
\cite{simplisma}, IPCA \cite{ipca} and SMAC \cite{smac}. Also, Band-Target
Entropy Minimization (BTEM) has been recently proposed which involves an
explicit (made by visual inspection) choice of spectral features (target
regions) to be retained in the course of constrained optimization \cite{btem}.
While efficient and highly flexible, supervised band selection has the
drawbacks of being neither fully automated, nor completely blind.

In order to cope with the non-negativity problem, several methods have been
proposed. In Refs.~\cite{plumbley,nmf,nnfica}, non-negativity is incorporated
in an approximate way already in the the basic decomposition algorithm. For a
discussion see Ref.~\cite{cichocki}. While the above are general purpose BSS
methods, there exist more specialized chemometrics (MCR) algorithms designed to
use non-negativity (see, {\it e.g.},
\cite{geladi,hopke,jiang,tauler,lavine,btem,dup}). In many of them, some
variant of principle component analysis is used to first prewhiten the data,
{\it i.e.} to minimize the off-diagonal elements of the covariance matrix. In a
class of algorithms known as MCR-ALS, prewhitening and the interdependence
minimization are done as in standard ICA, but non-negativity is then enforced
in a postprocessing step using the alternating least squares (ALS) technique
\cite{smd,mcrals,als}. In general, this gives better results than the
algorithms without ALS. But, since the interdependencies between the sources
are not checked during the ALS step, one runs the risk of worsening them. In
addition to that, applying {\it post facto} corrections to restore essential
properties which were present in original data, but which were ruined by the
main part of an algorithm, appears rather awkward conceptually.

Here we argue that a more straightforward strategy is not only possible, but
actually leads to de-mixing algorithms with notably improved performance. Our
approach is conceptually very simple and is based on mixture decomposition into
{\it strictly non-negative least dependent} components. Unlike other ICA
methods, it does not resort to prewhitening. Instead, we split the linear
de-mixing transformation into small random steps such that (i) on each step the
non-negativity is preserved {\it exactly}, and (ii) the dependencies between
the components tend to be reduced (the {\it mutual information} (MI) is
minimized non-greedily). Since greedy minimization would lead to trapping in
local minima of MI, we use a Metropolis-Hastings Monte Carlo strategy
\cite{metro,hast}. The algorithm (which we term SNICA, for Stochastic
Non-negative Independent Component Analysis) is explained in the next section.
The numerical results presented here show that our new method is
more efficient than previous MCR algorithms, including ICA and ICA-ALS type
methods \cite{smd} based on the same MI estimator.

\section{Method}
\label{meth}

As in most ICA and MCR approaches, we start out with the linear
mixture model

\begin{equation}\label{mix}
    {\bf X} = {\bf A} {\bf S},
\end{equation}

\noindent where ${\bf X}$ is the matrix of mixed signals (observed spectra),
${\bf S}$ are pure sources (spectra of individual mixture components), and
${\bf A}$ is the matrix of their concentrations (mixing matrix). We assume that
${\bf A}$ is a square $K\times K$ matrix, {\it i.e.} the number of sources $K$
is the same as the number of mixtures. ${\bf X}$ and ${\bf S}$ are of size
$K\times N$, where the number $N$ of frequencies in the spectra is much larger
than $K$. The task is to estimate blindly ${\bf S}$ and ${\bf A}$, given only
${\bf X}$. In our current setting the pure spectra and concentrations are
assumed to be non-negative and we seek a solution (an estimate for ${\bf S}$)
in the form of statistically least dependent components ${\bf Y}={\bf W}{\bf
X}$, with ${\bf W}^{-1}$ being an estimate for ${\bf A}$ up to scaling and
permutation of components (ambiguities inherent to the linear mixture model).
True and estimated mixing matrices can be compared using the Amari index
$P({\bf W}^{-1},{\bf A})$ which is a good measure of decomposition quality
\cite{bss,milca,smd}. The Amari index vanishes when the recovered
concentrations differ from the true ones only in scaling and permutation of
components, and it increases as the quality of decomposition becomes poor.
Thus, small \footnote{In practice, we find that good decomposition quality
roughly corresponds to Amari indices $P<0.05$, whereas $P>0.2$ generally characterizes
unacceptably poor performance.} values of the Amari index are
desirable.

We assume that each column vector ${\bf y}_i, \;i = 1,2, \ldots, N$ is a
realization of the same vector-valued random variable $Y=(Y_1, Y_2,\ldots,
Y_K)$, and we look for solutions where its $K$ components ({\it i.e.}, the
estimated individual spectra) have least mutual information. For a multivariate
continuous random variable with given marginal and joint densities $\mu_j(y_j)$
and $\mu(y_1,y_2,...,y_K)$, the mutual information is defined as

\begin{equation}\label{mi}
    I(Y_1,Y_2,...,Y_K)=\sum_{j=1}^K H(Y_j)-H(Y_1,Y_2,...,Y_K),
\end{equation}
\noindent where
\begin{equation}\label{hy}
    H(Y_j)=-\int \mu_j \ln \mu_j dy_j
\end{equation}
and
\begin{equation}\label{hym}
    H(Y_1,Y_2,...,Y_K)=-\int \mu \ln \mu \;dy_1 dy_2 ... dy_K
\end{equation}
are the differential entropies \cite{it}. Mutual information is zero if and only if
the $Y_j$ are strictly independent ({\it i.e.}, their joint density factorizes,
$\mu=\prod_j {\mu}_j$), and it is positive otherwise. Notice that the MI is
sensitive to all types of dependencies, not only linear correlations.

To assess the dependencies between the estimated spectra (given by
rows of ${\bf Y}$) we use the precise MI estimator based on nearest
neighbor statistics, denoted here as $I({\bf Y})$, developed in
Ref.~\cite{est}. It is derived from the Kozachenko-Leonenko estimate
for Shannon entropy \cite{koz}
\begin{equation}
   H(Y) = - \psi(k) + \psi(N) + \log c_D + {D\over N} \sum_{i=1}^N \log\epsilon(i)
   \label{KL}
\end{equation}
where $\psi(\cdot)$ is the digamma function, $\epsilon(i)$ is twice the
distance from the point ${\bf y}_i$ to its $k$-th neighbor, $D$ is the dimension of
$Y$, $c_D$ is the volume of the $D$-dimensional unit ball, and $N$ is the
length of sample. Mutual information could be computed by estimating the
entropies (\ref{hy}),(\ref{hym}) separately, with the same value of $k$, and
using Eq.~(\ref{mi}). But this would involve very different scales in the
marginal and joint spaces, leading to different biases. Looking
for nearest neighbors on the same scale in the joint $(Y_1,Y_2,...,Y_K)$ and
marginal $Y_j$ spaces significantly increases the chance that the biases in $H(Y_j)$
and $H(Y_1,Y_2,...,Y_K)$ cancel. This reduces the bias of the MI estimate.
Since Eq.~(\ref{KL}) holds for any value of $k$, one can first look for $k$ nearest
neighbors in the joint space and then use the hyper-rectangles defined by them
to calculate the number of neighbors in the marginal spaces. The estimate for MI
is then:
\begin{equation}
   I(Y_1,Y_2,...,Y_K) = \psi(k) - (K-1)/k - \langle \psi(n_1)+ \psi(n_2) + \ldots + \psi(n_K) \rangle + (K-1)\psi(N).
   \label{i2}
\end{equation}
where $n_j$ are the numbers of nearest neighbors in the spaces $Y_j, \;j = 1,2,
\ldots, K$; $\langle \ldots \rangle$ denotes the average over all points ${\bf
y}_i, \;i = 1,2, \ldots, N$. Implementation details are given in Ref.~\cite{est}.

Statistical errors in this MI estimator decrease with $k$, while systematic errors
increase. Empirically, we found that the decomposition performance is best
for the values of $k$ in the range 10-20, reflecting a balance between these
two types of errors. Also, spectral curve resolution by ICA
was shown to be more efficient if performed in the derivative space
\cite{smd,windig}, therefore in all simulations discussed below we estimate the
MI from the second derivatives of spectra with respect to frequency.

With this numerical measure of dependence, we can formulate the mixture
decomposition problem as minimization of $I({\bf Y})$ under the linear
de-mixing transformation ${\bf Y} = {\bf W}{\bf X}$, subject to the
non-negativity constraint \footnote {Note that the non-negativity of estimated
mixing matrix ${\bf W}^{-1}$ is not enforced explicitly, but it will be
nonetheless satisfied to high accuracy.} $y_{ik} \ge 0$.

In contrast to more traditional ICA methods that proceed by splitting the
de-mixing matrix into a prewhitening transformation (eliminating linear
correlations) and subsequent rotations \cite{ica,milca}, we approach the above
constrained optimization with a Monte Carlo strategy. To this end, the
de-mixing matrix ${\bf W}$ is factorized into small steps, ${\bf
W}=\lim\limits_{n\to\infty} {\bf W}_n$ with ${\bf W}_n =  {\bf M}_n {\bf
W}_{n-1}$, where each ${\bf M}_n$ is chosen randomly and is either a shear
transformation ${\bf T}$ in a two-dimensional (2D) subspace, or a rotation
${\bf R}$ in a 3D subspace (see below). The sequence of moves starts with ${\bf
Y}_0={\bf X}$, ${\bf W}_0={\bf I}$ (identity matrix). A move ${\bf Y}_n={\bf
M}_n {\bf Y}_{n-1}$ is immediately rejected, if it results in at least one
negative component $y_{n,ik} = ({\bf Y}_n)_{ik}$. Otherwise ({\it i.e.}, if all
$y_{n,ik}\ge 0$), it is accepted with probability one, if it leads to less
dependent components, $\triangle I = I({\bf Y}_n)-I({\bf Y}_{n-1})<0$, while it
is accepted with probability $\exp(-\triangle I/T)$, if $\triangle I >0$. Here,
$T$ is a fictitious temperature. This is the famous Metropolis-Hastings Markov
chain Monte Carlo method \cite{metro,hast}. In order to make the method more
greedy at later stages, when ${\bf W}_n$ is already close to the optimum, we
use a simple annealing schedule \cite{sa} where we start with high values of
$T$ and decrease it successively. Since both ${\bf R}$ and ${\bf T}$ matrices
act only on low-dimensional subspaces (and are the identity in the
complements), the difference $\triangle I$ in each step can be computed from
information in a 2D or 3D subspace only \cite{milca}. The convergence to the
global solution is monitored by computing the full dimensional mutual
information $I({\bf Y}_n)$, which guarantees that the least dependent, although
not necessarily independent, components are reached.

In our method, the mutual information is used to direct the search in the
global space, whereas the non-negativity constraint is important in that it
restricts solutions to physically meaningful subspace. This constraint has
strongest effect, if large parts of the constraint boundaries $y_k=0, \;k =
1,2, \ldots, K$, are populated by the pure sources, as then any linear
transformation is likely to violate non-negativity. This happens when each pure
spectrum has a high chance to be occasionally close to its zero baseline. Such
sources are called well-grounded \cite{plumbley} and are common in
spectroscopy.

The idea of using 2D shear transformations in tandem with the non-negativity
constraint was first proposed in the context of ``Positive Matrix Factorization"
\cite{pmf,pmf1}. The matrix ${\bf T}^{(ij)}(\alpha)$ representing a shear
$y_i' = y_i + \alpha y_j,\; y_j'=y_j$ in the $(i,j)$ subspace
is obtained from identity matrix by adding one off-diagonal element,
\begin{equation}\label{shearmat}
t^{(ij)}_{lm} = [{\bf T}^{(ij)}(\alpha)]_{lm}=  \delta_{lm} + \alpha \delta_{li} \delta_{mj},\; \;i \ne j \in
[1,K]; \;\; l,m=1,2,\ldots,K.
\end{equation}

\begin{figure}
\center
\includegraphics{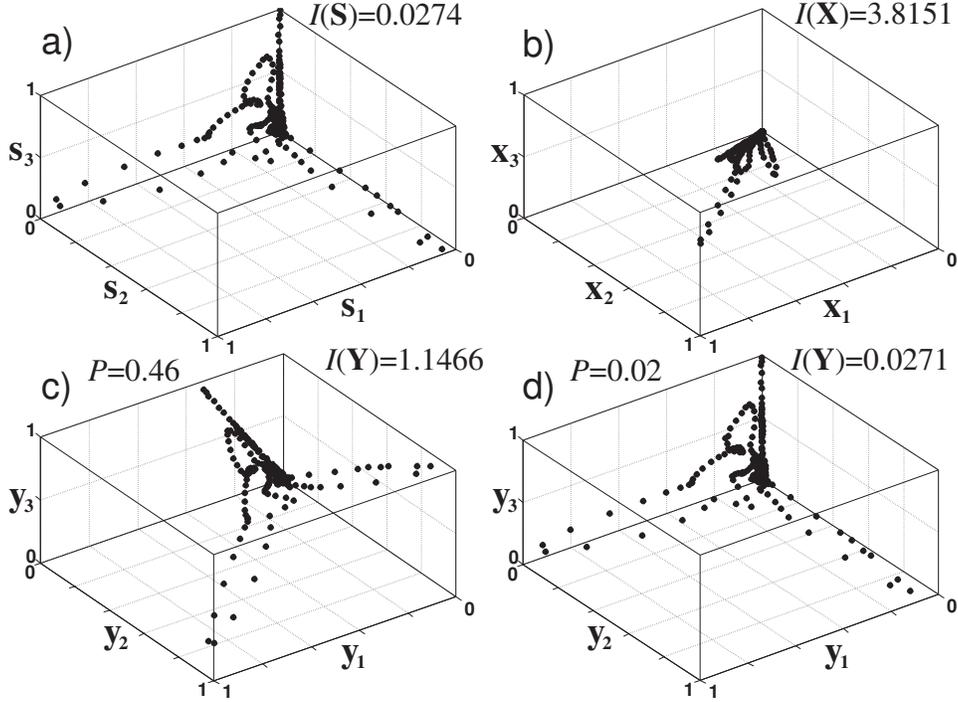}
\caption {De-mixing in three dimensions by shears and rotations. The scatter
plots are: pure sources (a); mixture signals (b); components (c) resolved by 2D
shears (trapped in a local minimum); components (d) recovered through
optimization with 2D shears and 3D rotations about the diagonal. Also shown are
the values of 3D mutual information ($I$) and the Amari index ($P$) of
incomplete (c) and successful (d) decompositions.} \label{trap}
\end{figure}

While 2D shear transformations are sufficient to decompose two-component
mixtures, they appear to be too restrictive to sample efficiently
higher dimensional de-mixing matrices. In higher dimensions, the minimization
through a sequence of shears can easily get trapped in local minima of the MI
landscape. One such case is illustrated in Fig.~\ref{trap} in which a
three-component mixture of exemplary infrared spectra is considered. First,
decomposition was attempted using only 2D shears, and Fig.~\ref{trap}c shows
the components found by constrained minimization. This configuration
corresponds to a local minimum of MI (compare to Fig.~\ref{trap}a and notice
the values of mutual information and very poor decomposition performance,
$P=0.46$). Since the components have reached the constraint boundaries and a
minimum of the cost function has been found to lie on this boundary, further
de-mixing shears can be accepted only by increasing $I({\bf Y})$, which can be
very slow. However, Fig.~\ref{trap}c suggests that the configuration needs to
be rotated in such a way that, afterwards, the three principal ``beams'' of
points could be brought close to the coordinate axes (and MI be decreased
further) by subsequent shears. This works indeed, and gives the correct pure
components with high precision, as seen in Fig.~\ref{trap}d.

The rotations needed to improve the efficiency are around the space diagonals
in randomly chosen 3D subspaces. Let us denote by ${\bf R}^{({ijk})}(\alpha)$
the matrix which performs a rotation by the angle $\alpha$ around the diagonal
in the positive octant of the subspace spanned by the components $(i,j,k)$.
It has the elements
\begin{eqnarray}
  r_{ii}^{({ijk})} =  r_{jj}^{({ijk})} =  r_{kk}^{({ijk})} &=& (1 + 2 \cos \alpha)/3, \nonumber\\
  r_{ij}^{({ijk})} = r_{jk}^{({ijk})} = r_{ki}^{({ijk})} &=& {1 \over 3}(1-\cos \alpha) + {\sin \alpha \over \sqrt 3}, \label{rotm2}\\
  r_{ik}^{({ijk})} = r_{ji}^{({ijk})} = r_{kj}^{({ijk})} &=& {1 \over 3}(1-\cos \alpha) - {\sin \alpha \over \sqrt
  3}. \nonumber
\end{eqnarray}
in the subspace spanned by $(i,j,k)$, while it is the unit matrix in the
complement, $r_{lm}^{({ijk})}=\delta_{lm}$ for $l,m \notin \{i,j,k\}$.
In the results shown below, random rotations and shears are used in ratio 1:1.

Of course, more sophisticated combinations of affine transformations
\cite{shr2,shr1} can be designed to sample the de-mixing matrix. This might
improve the performance of the algorithm further, especially in higher
dimensions. But it is not clear whether the added complexity would be
worthwhile.

As seen from Eqs.~(\ref{shearmat}),(\ref{rotm2}), the proposed moves ${\bf T}$ and
${\bf R}$ are parameterized by the ``angle'' $\alpha$. For each Monte Carlo
step its value is chosen randomly from some suitable range $\alpha \in
[-h, h]$. To realize an adaptive step-size control (where $h$ is adjusted
according to the progress of optimization), we use the following update rule

\begin{equation}\label{step}
 h_{n+1}=\left\{ \begin{array}{r@{\quad}l}
  f_1 h_n, & \mbox{if the move\,} {\bf M}_n \mbox{\,is accepted;}\\
  f_2 h_n, & \mbox{if it is rejected,} \end{array} \right.
\end{equation}

\noindent with the factors $f_1>1$, $f_2<1$ to be determined empirically. In our
simulations we took $f_1=1.06$, $f_2=0.98$ with initial $h_0=0.2-0.3$, although
the decomposition performance was found to be practically insensitive to
precise values of $h_0$ (see next section).

For the temperature annealing, we used rather simple schemes with only 2
or 3 cooling steps, with the temperature in the last step chosen close to zero.
This is because we found that the detailed annealing schedule
had no significant effect on the final results. It means, however, that the CPU times
needed in the following simulations might not be optimal. The initial value of
temperature $T_1$ was chosen of the order of the mutual information of pure
mixture components. Choosing $T_1$ too high leads to unfocused
search, while choosing it too small increases the risk to get trapped in false
minima. Both can result in poor performance, whereas our empirical choice of
$T_1$ ensured efficient decomposition in most cases.

At each temperature, the sequence of de-mixing moves is terminated when the
mutual information has reached its global minimum, according to some stopping
criterion. In our code we kept track of the minimum of $I({\bf Y}_n)$ during
the entire run,
\begin{equation}
I({\bf Y}_n)_{\rm min} = \min_{n'<n} I({\bf Y}_{n'}).
\end{equation}
Minimization is stopped when $I({\bf Y}_n)_{\rm min}$ does not decrease any
more during the last $M$ Monte Carlo steps, and the de-mixing matrix ${\bf
W}_{\rm min}$ that corresponds to $I({\bf Y}_n)_{\rm min}$ is given as output.
The chosen values of $M$ were some compromise between the speed and the quality
of decomposition.

With all the above together, the stochastic non-negative de-mixing procedure
with 2 cooling steps is then as follows:

\begin{enumerate}
    \item initialize $n=0$, ${\bf Y}_n={\bf X}$, ${\bf W}_n={\bf
    I}$; choose $h_n=h_0, T=T_1, M=M_1$;\\

\item \label{head} pick random $i,j,k \in [1,K]$, $\alpha \in [-h_n, h_n]$;
    \item  if $n$ is even:\\
     \hbox to 0.3cm{} compute shearing matrix ${\bf M}_n={\bf T}^{(ij)}(\alpha)$ (Eq.~(\ref{shearmat}));\\
    else:\\
     \hbox to 0.3cm{} compute rotation matrix ${\bf M}_n={\bf R}^{({ijk})}(\alpha)$ (Eq.~(\ref{rotm2}));

    \item propose the move ${\bf Z}={\bf M}_n {\bf Y}_n$, ${\bf V}={\bf M}_n {\bf
    W}_n$;

    \item if $\exists\; z_{lm}<0$:\\
     \hbox to 0.3cm{} reject the move, keep
          ${\bf Y}_{n+1}={\bf Y}_n$, ${\bf W}_{n+1}={\bf W}_n$ and go to (\ref{updt});

    \item estimate $\triangle I=I({\bf Z})-I({\bf Y}_n)$; pick random $p \in
    [0,1]$;

    \item if ($\triangle I<0$) or ($\triangle I>0$ and $e^{- \triangle I / T}>p$):\\
     \hbox to 0.3cm{} accept the move ${\bf Y}_{n+1}={\bf Z}$, ${\bf W}_{n+1}={\bf V}$;\\
     else:\\
     \hbox to 0.3cm{} reject the move, and keep ${\bf Y}_{n+1}={\bf Y}_n$, ${\bf W}_{n+1}={\bf W}_n$;

    \item if $I({\bf Y}_{n+1})$ reaches a new minimum:\\
     \hbox to 0.3cm{} store it together with ${\bf Y}_{\rm min}={\bf Y}_{n+1}$, ${\bf W}_{\rm min}={\bf W}_{n+1}$, and $n_{\rm min} = n+1$;

    \item \label{updt} adjust the step size $h_{n+1}$ according to
    Eq.~(\ref{step});

    \item $n=n+1$;\\
     if $n< n_{\rm min} + M$: \\
     \hbox to 0.3cm{}go to (2)\\
     else \\
     \hbox to 0.3cm{} if $T\neq T_2$: \\
     \hbox to 0.6cm{} choose $T=T_2$ and $M=M_2$, set $n = n_{\rm min}$, and go to (2)\\
     \hbox to 0.3cm{}else: \\
     \hbox to 0.3cm{}\hbox to 0.3cm{} output ${\bf Y}_{\rm min}$ and ${\bf W}_{\rm min}$ and stop.

\end{enumerate}

\section{Results and discussion}
\label{diss}

This section reports the results of extensive tests of the new algorithm
performed on a large sample of spectral resolution problems. Here we stick to
the strategy of statistical validation of decomposition performance developed
earlier \cite{milca,smd}. We stress that performing such tests on a
statistically representative set of precisely known mixtures is virtually the
only way to examine real efficiency of a de-mixing technique.

To compose a test set of randomized mixtures we first collected a pool of 99
experimental infrared absorption spectra in the range $550-3830$ cm$^{-1}$ (822
data points each) selected from the NIST database \cite{nist}. This pool of
chemical species was designed to contain organic compounds that have common
structural groups (halogen-, alkyl-, nitro-substituted benzene derivatives,
phenols, alkanes; alcohols, thiols, amines, esters) as well as a number of
``outliers'', {\it i.e.} structures dissimilar from all the rest. This was done
to be able to test how well the algorithm can treat both -- mixtures of highly
dependent sources \cite{smd} and cases in which very deep minima of MI are to
be located. After that, 4 sets with increasing dimensionality $K=3, 4, 7$ or
$10$ (6000 mixtures each) were constructed by randomly choosing $K$ normalized
pure spectra from the pool and applying random square mixing matrices
\footnote{Concentrations $a_{ij}$ were generated uniformly randomly in the
interval $[0,1]$.} of proper dimension $K$. Thus, our assessment of performance
is based on 24000 decompositions in total.

We examine the performance of the new decomposition technique (SNICA) in
comparison with the recently developed MILCA algorithm which uses the same MI
estimator \cite{milca}. In our previous study \cite{smd} MILCA was shown
to be very efficient when compared to other (MCR) decomposition algorithms on
typical spectral mixture problems (both synthetic and real). MILCA can be easily
combined with derivative preprocessing (to deal with overlapping spectra) and
ALS postprocessing (to improve non-negativity of estimates).
Decompositions by MILCA were performed in the second derivative space
\cite{smd}, and the postprocessing ALS iterations were stopped after 1000
steps (we verified that longer runs did not result in noticeable improvements).

Monte Carlo de-mixing was done with a two-step annealing scheme. The initial
temperatures and stopping criteria were $T_1=0.02,0.05,0.5,1$ and
$M_1=1000,1000,2500,8000$ (for $K=3, 4, 7,10$, respectively). For the
subsequent ``cooling'' stage these were set to $T_2=10^{-7}$ in all four cases
and $M_2=500,500,1500,3500$.

\begin{figure}
\center
\includegraphics{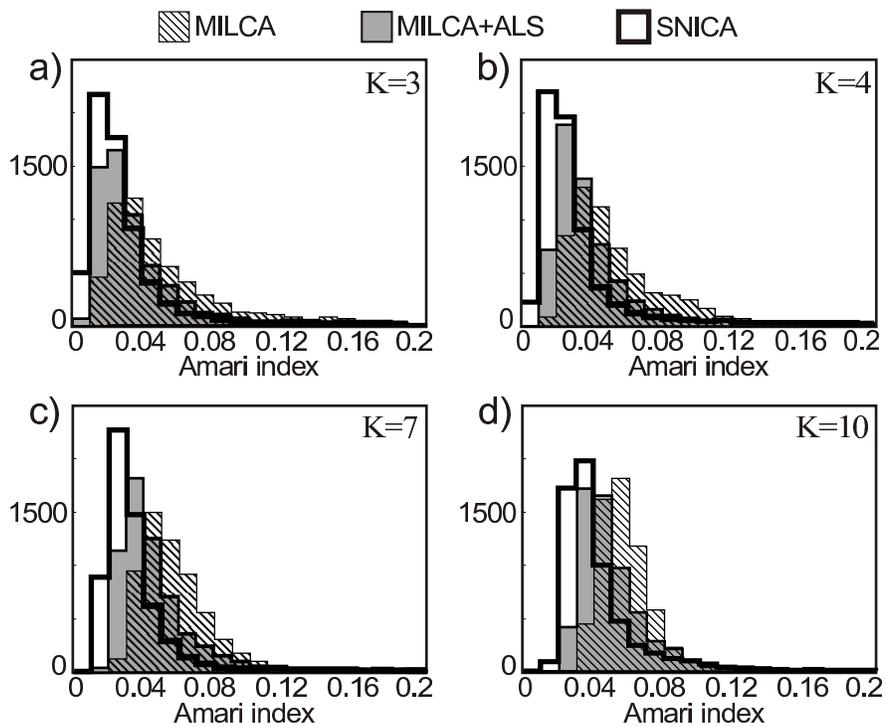}
\caption {Statistics of performance of MI-based algorithms (MILCA, MILCA+ALS
and SNICA) at decomposing 3-, 4-, 7-, and 10-component mixtures.} \label{distr}
\end{figure}

For each decomposition the Amari index $P({\bf W}^{-1},{\bf A})$ was computed.
Figure~\ref{distr} presents the distributions of the Amari indices with
increasing dimensionality $K$ of the mixture problem. Judging from the most
probable performance (peaks of the $P$ distributions) and the minimal achieved
values of $P$, SNICA clearly outperformed the other two algorithms. Also, the
Monte Carlo de-mixing seems to gain advantage as the number of mixture
components $K$ increases. In the current implementation (with a very simple
annealing scheme), SNICA is of order 10 times slower than MILCA and MILCA+ALS
for $K=3$, the algorithms have comparable speeds at $K=7$, and SNICA is on
average 2 times faster for $K=10$ (with our stopping criteria).

\begin{figure}
\center
\includegraphics{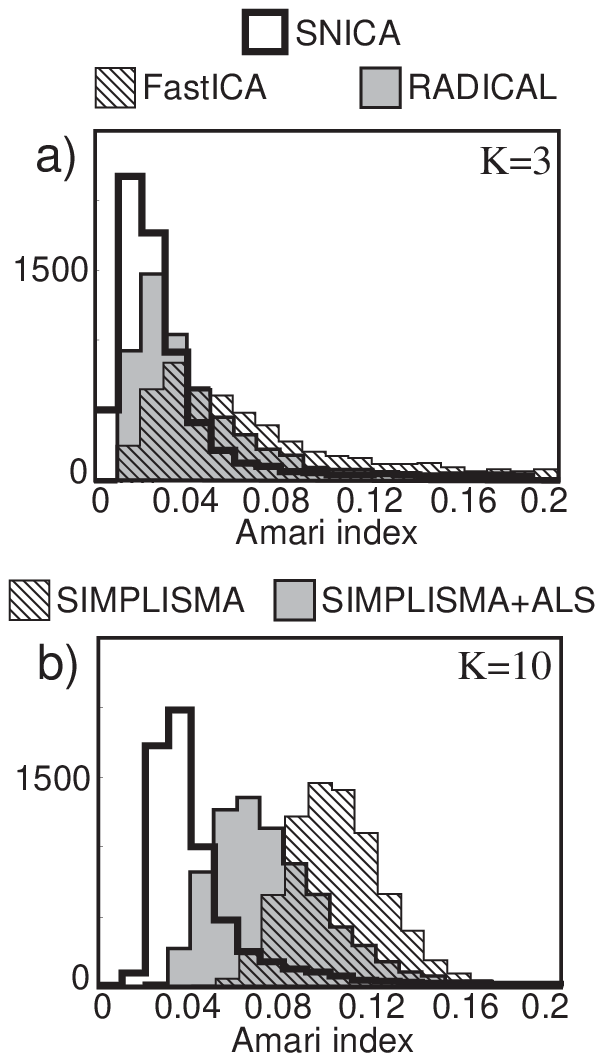}
\caption {Performance of SNICA in comparison to traditional ICA (a)
and MCR (b) algorithms (3- and 10-component mixtures, respectively).} \label{fastica}
\end{figure}

One of the mixture sets ($K=3$) was also processed \footnote{Second derivative spectra were used.
FastICA ran with its non-linearity parameter set to ``skew''. RADICAL worked without augmentation of data.}
by two representative ICA algorithms: the widely used FastICA \cite{fica} (for references on its
chemometrics applications see Ref.~\cite{smd}) and recently developed RADICAL
\cite{radical} (both algorithms rely on decorrelation of
data by PCA). The latter is similar to MILCA in many respects, notably in
speed and performance \cite{milca}, but uses a nearest neighbour estimator for entropies
$H(Y)$ \cite{vas}, not for MI.
We observe that the performance of SNICA is superior to these two ICA techniques
(Fig.~\ref{fastica}a). In particular, a significant fraction ($\sim 10\%$) of decompositions by
FastICA have $P>0.2$ (not shown), indicating relatively low performance.

A similar comparison was made with a state-of-the-art curve resolution algorithm SIMPLISMA
 \cite{simplisma} \footnote{We used the
SIMPLISMA code given in Ref.~\cite{windig}, choosing the ``offset'' parameter randomly
from the interval [0.01,0.1] which is about 1-10\% of the standard deviation of the
mixed signals. This is consistent with the choice made in Ref.~\cite{btem}.}
with and without ALS corrections (Fig.~\ref{fastica}b). We see that
ALS significantly improves SIMPLISMA decompositions, but there is still a remarkable gap
in performance between SIMPLISMA+ALS and SNICA. Fig.~\ref{fastica}b shows the results for
a high dimensional test set ($K=10$). We should mention, however, that the advantage of SNICA over
SIMPLISMA+ALS is less pronounced for smaller values of $K$, but it is still significant.

\begin{figure}
\center
\includegraphics{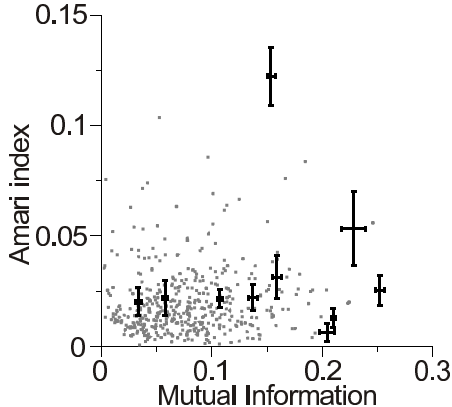}
\caption {500 SNICA decompositions ($K=3$) represented by the grey points in coordinates ($I({\bf S})$,$P$).
For 10 selected decompositions the one-standard-deviation error bars of estimated $I({\bf Y})$ and Amari indices $P$
were computed from 250-500 runs with randomized method parameters in each case (see text).} \label{std}
\end{figure}

Interestingly, decomposing mixtures of weakly and highly dependent sources by SNICA
turns out to be almost equally efficient (Fig.~\ref{std}, where the mutual information
scale spans the whole range of $I({\bf S})$ for triples of spectra in our test set).
This is in contrast to traditional ICA methods whose inaccuracies correlate with the
dependencies between pure sources, making decompositions of overlapping spectra
particularly difficult \cite{smd}.

In order to examine the sensitivity of the proposed technique to its parameters we
gathered the statistics of 10 decompositions with varied annealing temperature $T_1$,
initial Monte Carlo step size $h_0$ and stopping criteria $M_1$, $M_2$. Each decomposition was
attempted with these parameters taken uniformly randomly from their feasible ranges
for $K=3$: $T_1 \in [0.01,0.03]$, $h_0 \in [0.1,0.8]$, $M_1 \in [500,1500]$, $M_2 \in [0,1000]$
(the ``zero'' temperature at the second annealing step was kept fixed at $T_2=10^{-7}$).
As reflected by the error bars on Fig.~\ref{std}, the performance of SNICA as measured
by the Amari index is rather robust with respect to the parameter choices, suggesting
that Figs.~\ref{distr},\ref{fastica} are also robust.
The deviations from the mean final values
of the cost function $I({\bf Y})$ are also small for all decompositions indicating that
in all cases the global minima were reached. We note that these deviations may be simply
due to statistical errors in the MI estimation \cite{est}.

A separate set of high-statistics simulations with SNICA (for $K=3,4$) was performed, in which
the MI estimator of Ref.~\cite{est} was replaced by that of Ref.~\cite{dv} \footnote{We
used the implementation by P. Tichavsky available at {\tt
http://www.utia.cas.cz/user\_data/scientific/SI\_dept/Tichavsky.html}.} leaving
the rest of the SNICA code, all parameters and data unchanged. These estimators were
found to be comparable in speed and efficiency when used in ICA on spectral mixtures.
However, while the statistics of the Amari index (denoted $P_K$ and $P_D$) were
very similar for 3-component mixtures, it seems that the estimator of Ref.~\cite{est}
performs better in higher dimensions. Notably, in $70\%$ of cases the decompositions
of four-component mixtures using this estimator were better ($P_K<P_D$), with
the average values $\langle P_K \rangle=0.022$ and $\langle P_D \rangle=0.035$.

\begin{table}
  \centering
  \begin{tabular}{|l|c|c|c|c|}
  \hline
   & \begin{tabular}{c} SIMPLISMA \\ (+ ALS) \end{tabular} &
    BTEM & \begin{tabular}{c} MILCA\\ (+ ALS) \end{tabular} & SNICA  \\
  \hline
  {\small toluene}  & 0.971(0.973) & 0.954 & 0.987(0.994) & 0.973\\
  {\small $n$-hexane} & 0.994(0.995) & 0.992 & 0.990(0.991) & 0.981\\
  {\small acetone}  & 0.866(0.899) & 0.886 & 0.933(0.943) & 0.972\\
  {\small aldehyde} & 0.943(0.953) & 0.899 & 0.901(0.902) & 0.928\\
  33DMB  & 0.576(0.963) & 0.983 & 0.964(0.948) & 0.953\\
  DCM  &�0.969(0.967) & 0.904 & 0.909(0.966) & 0.964\\
  \hline
  {\small average}  & 0.887(0.958) & 0.936 & 0.947(0.957) & 0.962\\
  \hline
\end{tabular}
  \caption {Decomposition of 6-component experimental mixtures by MCR and ICA methods.
  The entries are normalized inner products between recovered and pure spectra of
  mixture constituents and their average values. Numerical data for SIMPLISMA and BTEM
  are from an independent study (Ref.~\cite{btem}, Table~3 which gives comparison with some
  other MCR methods -- IPCA \cite{ipca}, OPA-ALS \cite{opa}).} \label{tbl}
\end{table}

Now we turn to an experimental mixture problem to test SNICA in a realistic
application of spectral mixture decomposition. To facilitate comparison,
the example we chose was the one of Ref.~\cite{btem} where the performances of several
well-established curve resolution methods were compared. The experimental data
intended for blind separation consisted of FT-IR spectra ($950-3200$ cm$^{-1}$ at 5626 wavelengths)
of 14 mixtures of 6 solvents (toluene, $n$-hexane, acetone, 3-phenylpropionaldehyde
(aldehyde), 3,3-dimethylbut-1-ene (33DMB), and dichloromethane
(DCM)). In addition, the pure spectra of mixture constituents were measured
in separate experiments. Thus, this mixture problem offered a test
case complicated by experimental features typical to analytical
practice, such as non-linearities, instrumental noise, the presence
of spectrometer background and impurities (residual water and $CO_2$
in this case \cite{btem}).

First, we performed decomposition by
a PCA-based ICA algorithm (MILCA, with and without ALS
postprocessing) reducing the dimensionality of the data from 14 to 6
on the prewhitening step. The estimated components were then
compared to the pure sources, computing the normalized inner products
between them \cite{smd,btem}. The results are given in
Table~\ref{tbl} which also contains the respective values obtained
by two MCR methods -- SIMPLISMA \cite{simplisma} and BTEM
\cite{btem}. SNICA was applied to this data set with a 3 step
annealing schedule taking $T_1=1.5$, $T_2=0.1$, $T_3=10^{-7}$ and
$M=6000$ for all three steps. The Monte Carlo decomposition was done
in the full 14 dimensional space (using second-order Savitzky-Golay smoothing
differentiation \cite{sg} with a 5th order polynomial and a 51 point
window), thus 14 components were resolved. Then, we selected only 6
components which had the largest average contributions to the mixed
signals (this is similar to retaining only the components with
largest eigenvalues in the PCA dimension reduction). These 6 most
intense components were compared to the pure sources
(Table~\ref{tbl}). On this data set, the estimates produced by SNICA
are more accurate than those obtained by SIMPLISMA and BTEM without
ALS and compare favorably with the results obtained by MILCA and
ALS-versions of these algorithms. Notice that in order to produce
accurate estimates SIMPLISMA needs ALS postprocessing corrections \cite{dup,btem}
(see also Fig.~\ref{fastica}b). With these corrections, its performance seems to
match closely with SNICA, although higher statistics simulations (Fig.~\ref{fastica}b)
suggest that this might not be typical. Also, SNICA resolved all 6 components equally well, whereas the
other methods produced 2-3 relatively poorly estimated spectra
(notably, acetone and aldehyde appeared to be the most problematic due to
their highly overlapping spectra).

\section{Conclusions}
\label{conc}

In this paper we have proposed a new stochastic technique, SNICA, for blind
recovery of non-negative well-grounded sources from their linear mixtures.
Avoiding the PCA decorrelation as being counterproductive in certain cases, our
algorithm is based on a Metropolis Monte Carlo constrained minimization of
mutual information between recovered sources. The de-mixing transformation is
obtained as a sequence of random shears and rotations that resolves the least
dependent strictly non-negative components.

The SNICA approach can be related to some other PCA-free ICA methods (for
example, Infomax \cite{infomax} or TDSEP \cite{tdsep}) in that it does not
resort to the somewhat arbitrary decorrelation constraint \cite{card,naa} introduced
by PCA. Instead, it is replaced in SNICA by the non-negativity constraint which
reflects a natural property of spectral signals in the present application.
PCA simplifies and speeds up the decomposition considerably, but should be taken
with caution. If the sources are strongly correlated, then PCA produces
components which, in spite of being uncorrelated, are completely unphysical.
If an ICA method (consisting of a rotation of prewhitened components)
is used without ALS postprocessing, it has no chance to undo the errors made
by PCA prewhitening. For a specific example see Ref.~\cite{smd}. ALS postprocessing
improves the situation, but only partially.

Here we have tested SNICA in spectral curve resolution on a representative set
of mixture problems and experimental data. Our results indicate that the algorithm is superior in
performance to its PCA-based counterpart (MILCA, with the same mutual
information estimator) including the ALS-assisted version. We also found our
algorithm to outperform SIMPLISMA+ALS which represents a good reference point
among the state-of-the-art curve resolution chemometrics techniques. The method presented in this
paper performs blind decomposition, but the use of additional {\it a priori} information about
pure components (such as, {\it e.g.}, sparseness of spectral signals) might possibly improve the performance further.
In blind separation of multivariate time sequences, the perfomance of MILCA was
greatly enhanced if also mutual informations at different time lags were included in
the minimization \cite{milca}. In principle, the same can be done for SNICA,
but we have not yet made any tests to see how practical this is.

The assumptions of non-negativity and statistical independence of source signals
are met in a variety of practical problems in analytical spectroscopy and beyond.
The range of applications of blind
non-negative spectral mixture decomposition spans almost all quantitative
analytical approaches that rely on spectroscopic measurements and subsequent mixture analysis,
including, {\it e.g.,} hyperspectral remote imaging \cite{hyp, hyp1}, microscopy of
nano-structures \cite{opt} and biomedical samples \cite{smd,kra}, and separation
of independent light sources in multi-frequency astrophysical
observational data \cite{sky}. SNICA can be applied to all of them.

 The source codes of SNICA, MILCA and the MI estimator are freely available online at
{\tt http://www.fz-juelich.de/nic/cs/software}.

\section*{Acknowledgements}
\label{ack} We thank Prof. M.~Garland for providing us with the data set of
Ref.~\cite{btem}. S.~A. is grateful to Dr. Y.~Alaverdyan and Prof. S.~P.~Mushtakova
for discussions. We also thank the anonymous reviewers for their useful suggestions.

\end{document}